\documentstyle[fleqn]{tp}
\input psfig

\def\mpc{h^{-1}\mbox{Mpc}}
\begin{document}

\twocolumn[
\title{The 2dF QSO Redshift Survey}
\author{S.M. Croom$^1$, T. Shanks$^1$, B.J. Boyle$^2$, R.J. Smith$^3$,
L. Miller$^4$\\ and N.S. Loaring$^4$\\
{\it $^1$University of Durham, South Road, Durham, DH1 3LE, UK.}\\
{\it $^2$ AAO, PO Box 296, Epping, NSW 2121, Australia.}\\
{\it $^3$ MSSSO, Private Bag, Weston Creek, ACT 2611, Australia }\\
{\it $^4$University of Oxford, 1 Keble Road, Oxford, OX1, UK.}}
\vspace*{16pt}   

ABSTRACT.\
We present preliminary results from the 2-degree Field (2dF) QSO
Redshift Survey currently under way at the Anglo-Australian
Telescope.  This survey aims to determine the redshifts of $>25000$
QSOs over a redshift range of $0.3<z<3.0$ with the primary goal of
investigating large-scale structure in the Universe to high redshift
and at very large scales ($\sim1000\mpc$).

We describe the photometric procedure used to select QSO
candidates for spectroscopic observation.  We then describe results
from our first 2dF observations, which have so far
measured the redshifts for over 1000 QSOs.  We already find a
significant detection of clustering and have also found one close pair
of QSOs (separation $17''$) which are gravitational lens candidates.

To keep up to date with the current progress of the survey see: {\tt
http://msowww.anu.edu.au/$\sim$rsmith/QSO\_Survey/qso\_surv.html}
\endabstract]

\markboth{S.M. Croom et al.}{The 2dF QSO Redshift Survey}

\small

%%%% The star below makes the figure span both columns.  
%%%% Deleting the star makes the figure span only one column.  
%%%% At the moment, the command below is set up to leave 6cm
%%%% for the figure, but the figure itself is not included.  
%%%% To include the figure uncomment the line that starts with 
%%%% \centering, and comment out the \vspace*{6cm} line.  
%%%% There is nothing sacred about the 6cm height; you may 
%%%% prefer a different height for your figures.  
\begin{figure*}
\centering\mbox{\psfig{figure=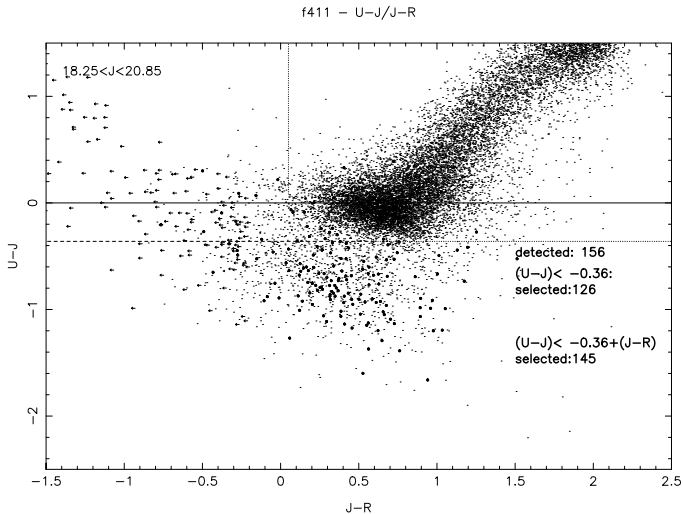,height=7cm}}
\caption[]{Multi-colour QSO selection for the 2dF QSO Redshift Survey 
in the F411 UKST field.  The small points indicate all stellar sources
in this field while the arrows show objects which only have upper
limits in the $R$ band.  The large points are known QSOs in this
field.  The dotted line is our final colour selection which selects
more QSOs than a simple $U-B_{\rm J}$ selection (dashed line).}
\label{colcol}
\end{figure*}

\section{Introduction and Aims}

The 2-degree Field Instrument\cite{2df95} at the AAT is a multi-object
fibre-fed spectrograph, allowing up to 400 spectra to be obtained
simultaneously.  This technological achievement has allowed a new
generation of redshift surveys to be undertaken which are an order of
magnitude greater in size than previous samples.  Previously, although
$\sim10000$ QSOs were known\cite{veron97} the largest homogeneous
catalogue only contained $\simeq1000$ QSOs\cite{hewett95}.

The 2dF QSO Redshift Survey aims to measure the redshifts of over
25000 QSOs selected in a homogeneous manner via $UB_{\rm J}R$
multi-colour selection (see Section \ref{mulcol}).  The sample will
span the redshift range $0.3<z<3.0$ and will be $>90\%$ complete at
$z<2.2$.  The survey covers an area of 740 deg$^2$ in two declination
strips which are each $75^{\circ}\times5^{\circ}$, one strip is at
$\delta=-30^{\circ}$ near the South Galactic Pole while the second is
at $\delta=0^{\circ}$ in the North Galactic Cap.

The main scientific goals of the QSO survey are the following:
\begin{enumerate}
\item To measure the QSO power spectrum from small scales $\sim2\mpc$
to the large scales probed by {\it COBE}, $>1000\mpc$.  This will also
allow us to place limits on $\Omega_{\rm baryon}$ by searching for
acoustic oscillation features ( or ``doppler peaks'') present in the
power spectrum when $\Omega_{\rm baryon}$ is a significant fraction of
$\Omega_{\rm CDM}$.
\item Estimate cosmological parameters, in particular $\lambda_0$,
the cosmological constant, from geometric distortions in the
clustering of QSOs at high redshift.
\item Trace the evolution of QSO clustering from $z=0.3$ to $z=3$ to
test models of structure formation and evolution in the Universe.
\end{enumerate}
There are of course many other studies which can be carried out on a
sample of this size, including: accurate determination of the QSO
luminosity function to $z=2.2$; investigation of the radio properties
of QSOs; the study of broad-absorption-line QSOs; and searches for
gravitational lenses.  The catalogue will also provide a database of
objects to be followed up, for example, with high resolution
spectroscopy to find ${\rm Ly}\alpha$ and metal absorption line systems.

\section{Multi-colour Selection of QSO Candidates}\label{mulcol}

The QSO candidate catalogue for spectroscopic observation is based on
broad band $UB_{\rm J}R$ photometry from APM measurements of UKST
photographic plates.  The survey comprises 30 UKST fields, and in each
of these fields one $B_{\rm J}$ plate, one $R$ plate and up to 4 $U$
plates/films are used.  In each UKST field area we have obtained CCD photometry
from the 0.9m Telescope at CTIO and the INT and JKT on La Palma in
order to independently calibrate the photographic
magnitudes\cite{bsc95,crpsbs98}.  We also correct for the 
vignetting/variable desensitization in the photographic plates
\cite{smith98}.  QSOs candidates are then selected by their blue
colours in the $U-B_{\rm J}$ vs. $B_{\rm J}-R$ plane.  An example of
this is shown in Figure \ref{colcol} for the F411 UKST field.  This
shows that our multi-colour selection process is more effective than a
simple $U-B_{\rm J}$ selection and based on previously known QSOs
(discovered by a variety of methods) in our survey area we expect to
be 93\% complete\cite{croom97}.  The incompleteness is due scatter
from both photometric errors and variability (when the photographic
plates do not share similar epochs).

The magnitude limits of the survey are $18.25<B_{\rm J}<20.85$ in the
case of the 2dF observations.  In order to provide a well determined
luminosity function at bright magnitudes we observe brighter QSO
candidates, $17.0<B_{\rm J}<18.25$, using FLAIR on the UKST.

\section{Results from the First 1500 QSOs}

\begin{figure*}
\centering\mbox{\psfig{figure=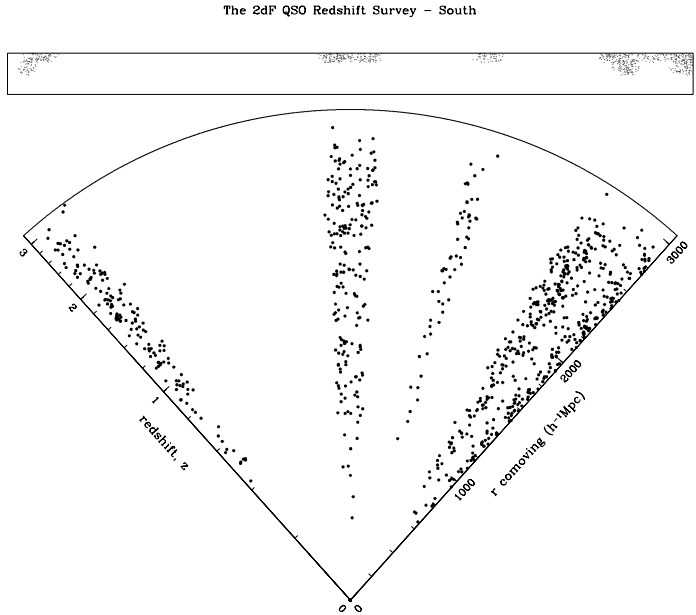,height=8.5cm}}
\caption[]{The positions of observed QSOs in the SGP survey region
($\delta=-30^{\circ}$).  This area currently contains 910 QSOs.  The
rectangle at the top of the plot will eventually be completely filled
with QSOs.}
\label{wedge}
\end{figure*}

We have currently obtained redshifts for just over 1500 QSOs from a
variety of sources.  Over 1300 of these are from 2dF with the
remainder from FLAIR observations of bright QSOs, Keck observations
of radio-loud QSOs and observations of close pairs of QSOs from the
2.3m ANU Telescope.  The 2dF observations are carried out in
collaboration with the 2dF Galaxy Redshift Survey (see Maddox, this
volume) as the QSO survey covers a subset of the galaxy survey area,
so the two catalogues have been merged to increase the efficency of
observations.  The data are reduced using pipeline reduction software
and QSO identification and redshift determination is
carried out using software specifically designed for this task.
Spectroscopy confirms that $\sim53\%$ of our
candidates are positively identified as QSOs with the contamination
consisting mainly of galactic subdwarfs and compact blue galaxies.

\begin{figure}
\centering\mbox{\psfig{figure=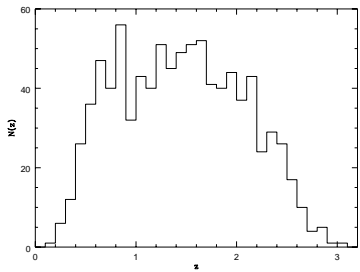,width=5cm}}
\caption[]{The number-redshift histogram for the 2dF QSOs with
currently measured redshifts.}
\label{nz}
\end{figure}

\begin{figure*}
\centering\mbox{\psfig{figure=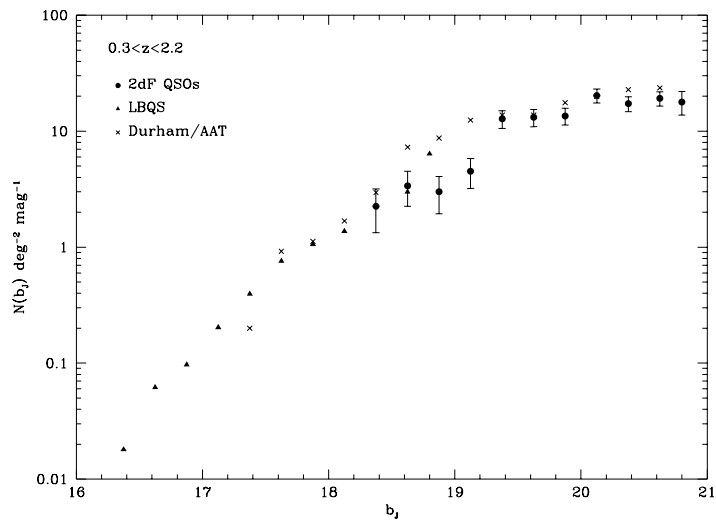,width=10cm}}
\caption[]{The number-magnitude relation for 2dF QSOs (filled
circles), compared to the LBQS (triangles) and Durham/AAT (crosses)
surveys.} 
\label{nm}
\end{figure*}

\begin{figure*}
\centering\mbox{\psfig{figure=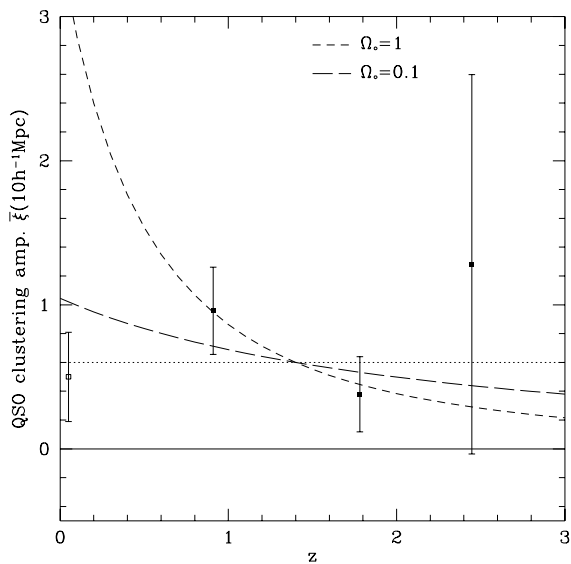,height=8cm}}
\caption[]{Evolution of QSO clustering as measured by the 2dF QSO
Survey combined with the Durham/AAT, CFHT and LBQS samples (filled
squares).  The low redshift result (open square) is a combination of the
EMSS\cite{bm93} and IRAS\cite{gs94} AGN correlation functions.  Two
simple linear theory models are 
shown (for $\Omega_0=1$ and $0.1$) as a comparison.}
\label{evol}
\end{figure*}

\begin{figure*}
\centering\mbox{\psfig{figure=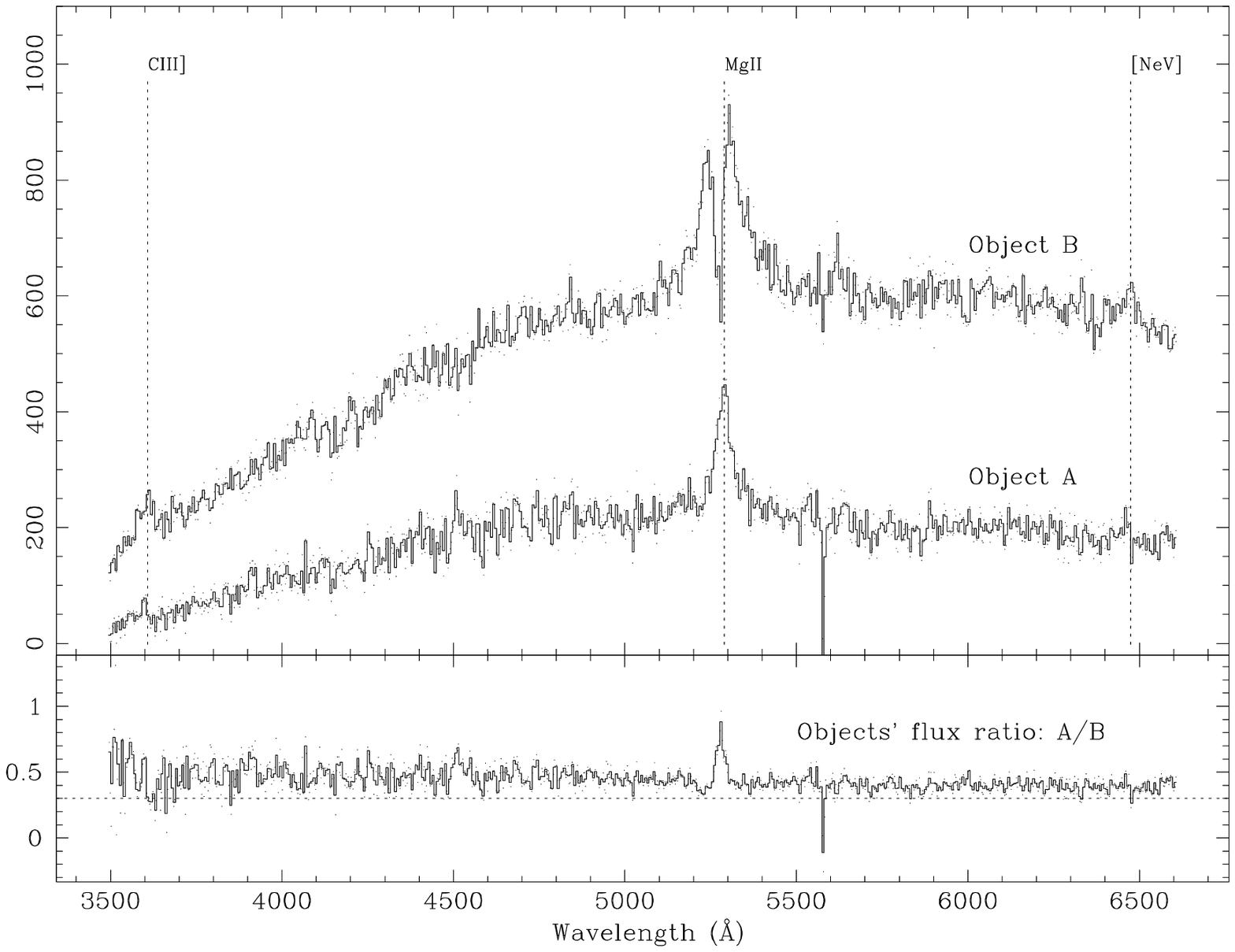,height=7.0cm}}
\caption[]{Spectra of the close pair of QSOs at $z=0.89$.  The spectra
were obtained using the RGO spectrograph at the AAT.  Below the two
spectra we show the flux ratio of the two, which is close to constant
apart from absorption in the MgII line of object B.}
\label{pair}
\end{figure*}

Although the survey is still only $\sim4\%$ complete, Figure
\ref{wedge} demonstrates the eventual scale of the survey.  This plot
shows the positions of all the QSOs with redshifts measured in the
$\delta=-30^{\circ}$ slice, which currently shows 910 QSOs, but
will eventually contain in excess of $\sim12500$.  The redshift
distribution of the QSOs is shown in Figure \ref{nz}.  The
multi-colour selection allows us to select QSOs at $z>2.2$ although
the survey is not highly complete above this redshift. The
number-magnitude relation is shown in Figure \ref{nm} compared to the
LBQS\cite{hewett95} and Durham/AAT\cite{bfsp90} surveys.  The 2dF QSO
Survey is marginally lower 
than the Durham/AAT survey, consistent with our predicted
completeness, however this slight incompleteness is compensated for by
our much reduced contamination.

Although the area coverage of the survey is currently disjoint, we can
make provisional estimates of the clustering.  We use the integrated
correlation function $\bar{\xi}$ out to a scale of $10\mpc$ as a
measurement of clustering.  For QSOs with $0.3<z<2.2$ we find
$\bar{\xi}(10)=0.55\pm0.27$ from 929 QSOs.  This is consistent with
other measures, e.g. $\bar{\xi}(10)=0.83\pm0.29$\cite{cs96} from the
Durham/AAT\cite{bfsp90}, LBQS\cite{hewett95} and CFHT\cite{crampton89}
samples.  We plot the clustering amplitude as a function of redshift
in Figure \ref{evol}, here we combine the 2dF data with the data
discussed above.  We see a marginal decrease in clustering from
$z\sim0.9$ to $z\sim1.8$.  Of course the errors in the
final survey will be more than $5\times$ smaller than the errors shown in
this plot.

We have also begun a long-slit spectroscopic program to obtain
redshifts for all the close pairs of QSO candidates
($\Delta\theta<20''$) which cannot be
simultaneously measured with 2dF (due to a limit on the minimum
fibre separation).  Such pairs are potentially some of the most
interesting objects in our catalogue as they are candidate wide-angle
gravitational lenses or binary QSOs.  With observations of $\sim25\%$
of the close pairs complete, we have found 8 QSO pairs, but only one
pair of QSOs with identical redshifts.  This pair has a separation of
$17''$ and is at a redshift of $z=0.89$.  Spectra of these two objects
are shown in Figure \ref{pair}.  The number of these lens can be used
to constrain $\lambda_0$.

\section{Conclusions: A Survey of 25000 QSOs}

The 2dF instrument allows the creation of a QSO survey some 25 times
bigger than the previous largest sample.  With spectroscopic
observations only $\sim4\%$ complete we have already discovered
$>1000$ QSOs.  QSO clustering measurments already show significant
detections of clustering, and possibly evolution in clustering.
With the increasing efficiency of 2dF the rate of redshift measurement
is increasing dramatically and the expected completion date of the
survey is the end of 2000.

%%%% The star below makes the section heading appear without 
%%%% a section number.  

%%%% If you have more than 99 references, but fewer than 1000, 
%%%% then you need to change 99 to 999.  

\end{document}